\documentclass{article}
\usepackage[utf8]{inputenc}
\usepackage{jcappub}
\usepackage[dvipsnames]{xcolor}
\usepackage{amssymb,amsmath}
\usepackage{subcaption}
\usepackage{graphicx}

\newcommand{\HI}{\mathrm{HI}}

\newcommand{\eq}[1]{Eq.~(\ref{#1})}

\newcommand{\fig}[1]{Figure~\ref{#1}}

\def\ie{{\em i.e.}~}
\def\eg{{{\em e.g.}~}}

\def\kMpc{\, h \, {\rm Mpc}^{-1}}

\author[a,b]{Emanuele Castorina,}
\author[a,b]{Martin White,}

\affiliation[a]{Department of Physics, University of California, Berkeley, CA 94720}
\affiliation[b]{Berkeley Center for Cosmological Physics, Berkeley, CA 94720}

\emailAdd{ecastorina@berkeley.edu}
\emailAdd{mwhite@berkeley.edu}

\title{Measuring the growth of structure with intensity mapping surveys}
\date{January 2019}

\abstract{Line intensity mapping offers a new avenue for constraining cosmological parameters in the high redshift Universe. However measurements of the growth of structure, a sensitive probe of gravity, are affected by a well known degeneracy with astrophysical parameters, encoded in the mean brightness temperature of the specific line. In this work we show how to break this degeneracy, to a level that could allow constraints of the amplitude of cosmological fluctuations at the percent level, using information in the mildly non-linear regime of structure formation as described by Lagrangian Perturbation Theory. We focus on the 21-cm line with forecasts for HIRAX and the proposed Stage {\sc ii} experiment as illustrations.}

\begin{document}
\maketitle
\flushbottom

\section{Introduction}

Line intensity mapping has recently emerged as a potentially powerful technique to constraint cosmological parameters. In particular the 21-cm transition of the hydrogen atom can be used to probe the distribution of the neutral hydrogen, in the post-reionization era, using receivers operating at radio wavelengths.
In total intensity, the 21-cm emission has been detected by cross-correlating the 21-cm data collected at the Green Bank Telescope with galaxy positions in Deep-II \cite{Chang} and 2DF \cite{Masui}, and more recently by Parkes radio telscope with Wiggle-z\cite{Parkes}.
Several new instruments, such as CHIME \cite{CHIME}, HIRAX \cite{HIRAX}, BINGO \cite{Bingo}, Tianlai \cite{Tianlai} and the SKA\cite{SKAcosmo18}, plan to measure the 21-cm signal in auto-correlation over large area of the sky in the redshift range $1<z<6$.
Recently, within the Cosmic Visions: ``Dark Energy'' program, a new, 21-cm interferometer has been proposed to measure large-scale structure at high redshift ($2<z<6$). 

One of the primary goal of the aforementioned 21-cm surveys it to detect the Baryon Acoustic Oscillations and therefore measure the geometry and expansion rate of the Universe across cosmic time.
Traditional galaxy surveys usually have another science goal, to measure the growth rate of Large Scale Structure (LSS) using redshift space distortions (RSD). The latter in particular is a very sensitive probe of gravity \cite{Weinberg13}.
In line intensity mapping, the signal we measure is the product of the cosmological fluctuations in the specific line times the mean brightness temperature of the line, $\bar{T}_b(z)$, which is proportional to the total luminosity, or mass, of the sources emitting the photons. 
This implies that in order to measure the growth of structure using intensity mapping we need to break the degeneracy between $\bar{T}_b(z)$ and the amplitude of cosmological fluctuations $\sigma_8(z)$, which in linear theory would be perfectly degenerate\footnote{The BAO measurements are instead not affect by this degeneracy, as the constraints are basically independent from an overall rescaling of the power spectrum.}. This fact has been previously overlooked in the literature \cite{Bull2015,SKARSD,Pourtsidou17,SKAcosmo18}, where linear theory has been used to forecast the constraining power of future 21-cm experiment under the (unrealistic) assumption that $\bar{T}_b$ was perfectly known.

It has also been recently argued that the $\bar{T}_b$ degeneracy can be broken by combining the HI dataset with other probes \cite{Obuljen18,Chen19}.  
The alternative route we take is to model the signal using beyond linear scales.  
The purpose of this short note is therefore two-fold.  First we wish to highlight how mildly non-linear clustering allows us to break the $\bar{T}_b$-$\sigma_8(z)$ degeneracy.  The differences from linear theory appear at scales large enough that perturbation theory with a parameterized bias model is still quantitatively reliable.  Second we advocate for the use of the Lagrangian Perturbation Theory for biased tracers as a means of forecasting parameter constraints for high-redshift observations.  
In this work we will focus on the 21-cm line, but our results apply more broadly to any other emission line with unknown mean brightness, see \cite{Kovetz17} for a comprehensive list of intensity mapping surveys.

This paper is organized as follows. In Section \ref{sec:signals}, we describe the signal and noise model for HIRAX and the proposed Stage {\sc ii} survey, and the assumptions that go into our modeling.
The forecasting methodology is described in Section~\ref{sec:fisher} and largely follows Ref.~\cite{Chen19}.
Finally, Section~\ref{sec:results} describes our results and our conclusions are summarized in Section~\ref{sec:conclusions}.
For numerical results we assume a spatially flat $\Lambda$CDM model with $\Omega_m\simeq 0.31$ and $\sigma_8\simeq 0.8$, consistent with recent measurements \cite{Planck18-I}.

\section{The 21-cm power spectrum model}
\label{sec:signals}
\subsection{Signals}
We shall largely follow Ref.~\cite{Chen19} in modeling the 21-cm signal and the HIRAX and Stage {\sc ii} instruments.  We assume most of the hydrogen in the Universe is ionized, and the 21-cm signal comes only from self-shielded regions such as galaxies (specifically between the outskirts of disks until where the gas becomes molecular within star-forming regions) \cite{VN18}.  We assume the HI is a biased tracer of the matter field, in redshift space, though we do not require that the bias be linear or scale-independent.  Typical values of the bias on large scales are $b_{HI}=2-6$ over the range $2<z<6$ \cite{Castorina17}.

The 21-cm signal is proportional to $\bar{T}_b$, which is the mean brightness temperature, related to the mean intensity at frequency $\nu$ as $I_\nu=2k_B\bar{T}_b(\nu/c)^2=2k_B\bar{T}_b/\lambda^2$.  The mean, $\bar{T}_b$, is proportional to $\Omega_{\rm HI}$, the HI density in units of the critical density \cite{Field1959}.
Unfortunately the value of $\Omega_{\HI}$ is quite uncertain (see e.g.~Refs.~\cite{Padmanabhan15,Crighton15} for recent compilations of data).

We model the clustering of the HI using an `effective field theory' (EFT) version of the redshift-space Zeldovich power spectrum \cite{Zel70,Por14,VWA15}, including first and second order Lagrangian bias ($b_1$ and $b_2$) following Refs.~\cite{Mat08,CLPT,Whi14}. This model was shown in Ref.~\cite{VN18} to provide a very good description of the real space HI power spectrum measured in hydrodynamical simulations.
The multipoles of the power spectrum can be expressed as
\begin{align}
    P_{\rm HI,\ell}(k) &= \left[ \left(1+\alpha_\ell k^2\right)P_{Z,\ell}(k) + b_1 P_{b_1,\ell}(k)
    + b_2 P_{b_2,\ell}(k) \right. \nonumber \\
    & \left. + b_1^2P_{b_1^2,\ell}(k) + b_2^2 P_{b_2^2,\ell}(k) + b_1b_2P_{b_1b_2,\ell}(k) \right]
\label{eqn:Pk_ZEFT}
\end{align}
where $b_1$ and $b_2$ are (Lagrangian) bias terms and $\alpha_\ell$ represent the lowest-order EFT terms (assumed to go as $\sigma_8^4$).
There are several routes to the redshift-space power spectrum \cite{Vlah19} in Lagrangian Perturbation Theory.  We choose to explicitly Hankel transform the correlation function multipoles, though we have checked that we obtain the same answer using the code of ref.~\cite{VCW16,Modi17}.  The explicit expressions for the power spectrum terms in Lagrangian perturbation theory can be found in several places in the literature, see for instance refs.~\cite{Mat08,Whi14,VCW16,Modi17}.

While it is an approximation to include the non-linear corrections from the Zeldovich expression and the counter terms but neglect 1-loop terms, we find that at high $z$ and low $k$ this approximation is actually numerically quite accurate in comparison to N-body simulations and the resulting expressions are much simpler to evaluate than the full 1-loop theory.  The signal is smooth in $\mu$, so we include only $\ell=0$, 2 and 4 in our calculation of $P_{\rm HI}(k,\mu)$.

\begin{figure}
    \centering
    \resizebox{.325\columnwidth}{!}{\includegraphics{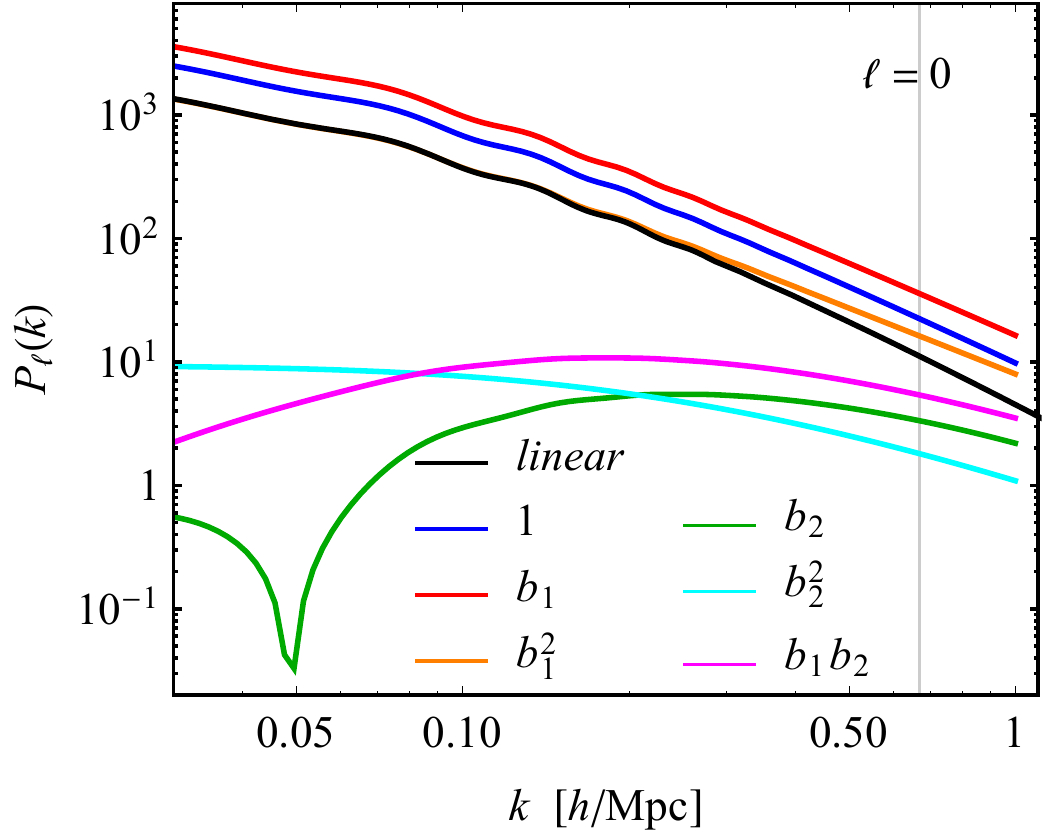}}
    \resizebox{.325\columnwidth}{!}{\includegraphics{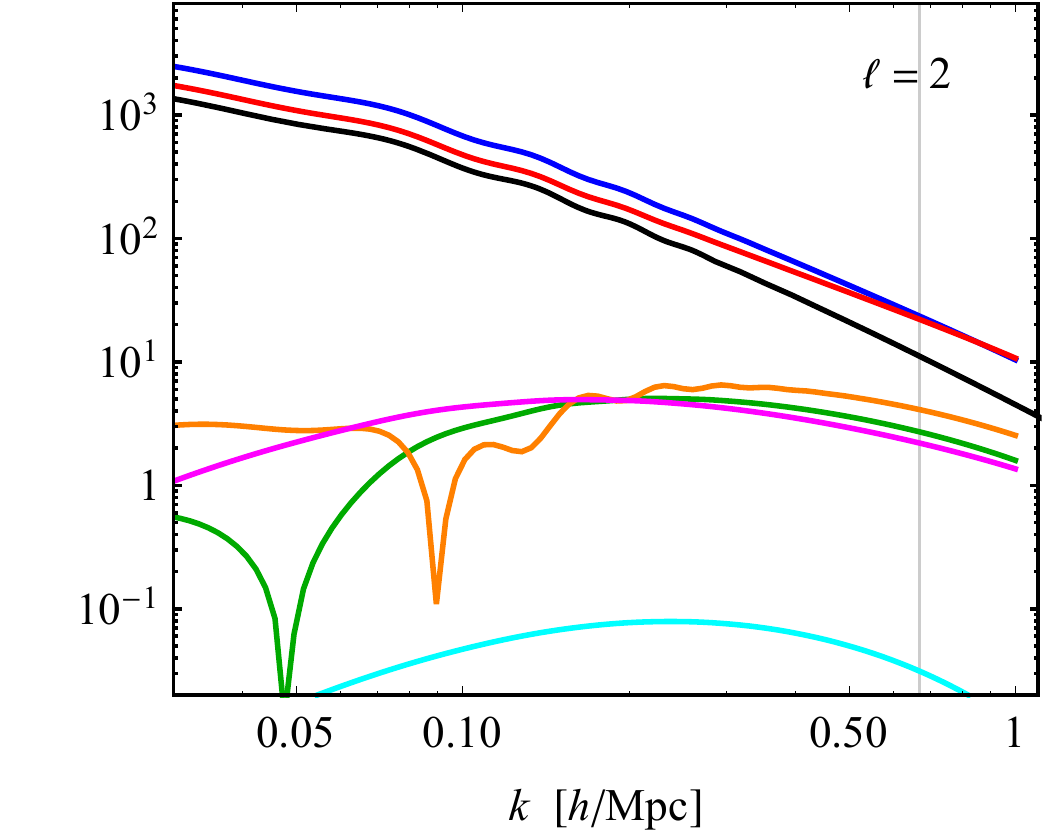}}
    \resizebox{.325\columnwidth}{!}{\includegraphics{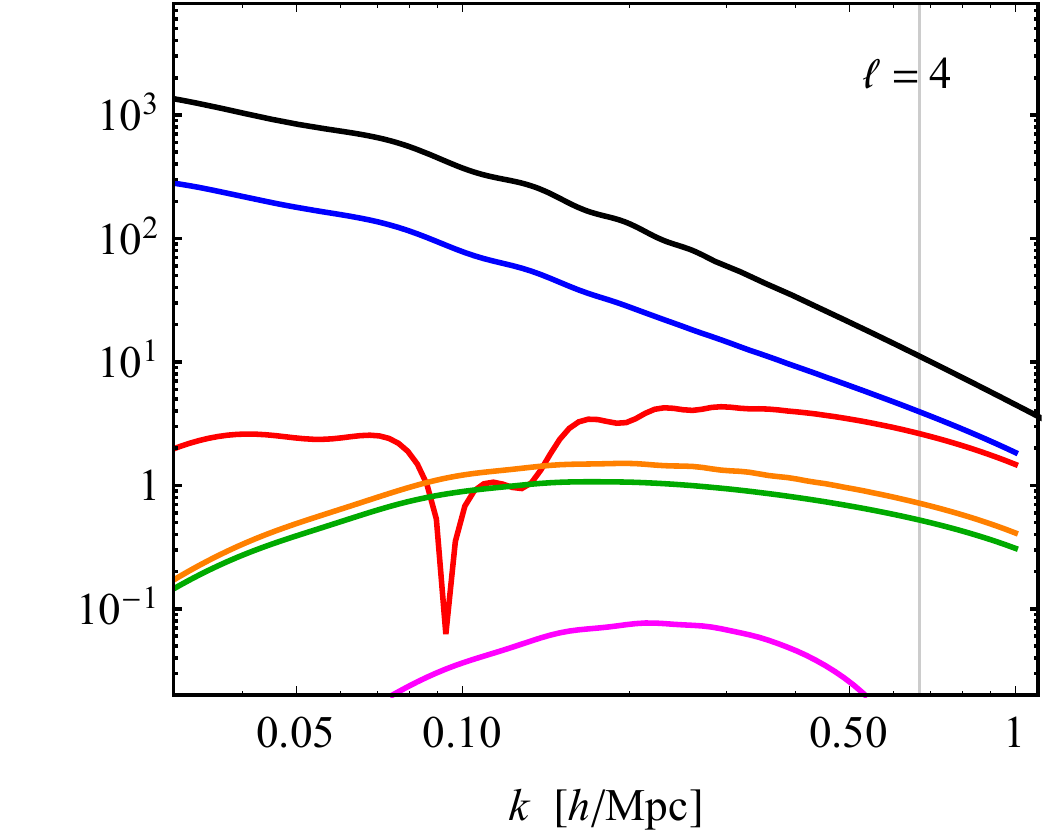}}
    \caption{The contributions to the monopole (left), quadrupole (middle) and hexadecapole (right) moments of the redshift-space power spectrum at $z=4$, see Eq.~(\ref{eqn:Pk_ZEFT}).  We highlight $z=4$ as it is the middle of the range probed by a Stage {\sc ii} instrument, but the contributions at the other redshifts look qualitatively similar.  We have not shown the $\alpha$ terms as they are simply proportional to $k^2$ times the term labeled `1'.  The grey, vertical, dotted line marks the non-linear scale (see text).}
    \label{fig:pieces}
\end{figure}

In Fig.~\ref{fig:pieces} we show the contributions to the three lower multipoles of the power spectrum at $z=4$. The black line in each panel displays the corresponding linear theory dark matter power spectrum, whereas the different terms in \eq{eqn:Pk_ZEFT} are shown with coloured lines. On large scales the dominant contribution to the Zeldovich power spectrum is coming from the terms proportional to the linear bias $b_1$, whose scale dependence is very similar to the linear theory power spectrum. For $k>0.1\text{-}0.2 \kMpc$ we start seeing differences between linear theory and our model, something will allow us, in the next section, to break the degeneracy between the brightness temperature and cosmological parameters. It is also worth noticing that the $b_1$ terms and the `1' term deviate from each other at approximately the same scale, where non-linear bias terms also become important.

The accuracy to which we can disentangle the brightness temperature from $\sigma_8(z)$ obviously depends on the fiducial value of the bias parameters.
In order to fix reasonable values for the latter, we fit the redshift-space clustering to a mock HI sample derived from a high resolution N-body simulation.  The simulation used a TreePM code \cite{TreePM} to evolve $2560^3$ particles in a $256\,h^{-1}$Mpc box assuming a $\Lambda$CDM cosmology with $\Omega_m\simeq 0.31$ and $h\simeq 0.68$.  This is the same simulation as used in Refs.~\cite{Stark15a,Stark15b} and we refer the reader to those papers for more information.
We populated the halos and subhalos in that simulation with HI following Ref.~\cite{Castorina17} and measured the multipole moments of the redshift-space power spectrum at $z\simeq 2-6$ (see also Fig.~8 of Ref.~\cite{CVDE-21cm}).  The $\ell=4$ moment is too noisy to be useful, but $\ell=0$ and $2$ are resolved.  We then fit $b_1$, $b_2$ and $\alpha_\ell$ to the $\ell=0$ and $2$ spectra, finding good fits for $k<k_{\rm nl}$.  We set $\alpha_4=0$.  The values of $b_1$, $b_2$, etc.~are consistent with expectation from Ref.~\cite{Chen19}.
Our main conclusions will be independent of the precise details, but should also be regarded as tentative due to the extreme uncertainty in the manner in which HI traces large-scale structure at such high redshifts.

\subsection{Noise}

In an interferometer the fundamental datum is the correlation between two feeds (or antennae) $i$ and $j$, known as a visibility \cite{TMS17}.  On the scales of interest to us, and for an intensity measurement, the visibility measures essentially the Fourier transform of the sky emission at a wavenumber set by the spacing $\vec{u}_{ij}$ of the two feeds (in units of the observing wavelength). In particular, such feeds correspond to a comoving wavenumber with component perpendicular to the line-of-sight $k_\perp = 2\pi \vec{u}_{ij}/ \chi(z)$. 
In interferometry the angular resolution of the survey is roughly set by the \emph{longest} baseline. As we will show momentarily this implies, for both HIRAX and Stage {\sc ii}, that one loses signal in the perpendicular direction on scales much smaller than what can be achieved with perturbation theory. This is in contrast to single dish surveys, like SKA-MID, that have low angular resolution and cannot measure high $k_\perp$ modes \cite{VN17}. For both types of instruments the radial resolution is given by the frequency channel bandpass, which can in principle be made as small as required.
The visibility noise is inversely proportional to the number (density) of baselines, $n(\vec{u})$, normalized such that $\int n(u)d^2u=N_{\rm dish}(N_{\rm dish}-1)/2$.  It is explicitly given by \cite{ZalFurHer04,McQ06,Seo2010,Bull2015,SeoHir16,Wol17,Alonso17,White17,Obuljen18,Chen19}
\begin{equation}
  P_{th} = T_{\rm sys}^2
    \left( \frac{\lambda^2}{A_{\rm e}} \right)^2
    \left( \frac{4 \pi f_{\rm sky}}{\Omega_p(z)} \right) \frac{1}{n_{\rm pol}\nu_0 t_{\rm obs} n(\vec{u})}
    \frac{d^2V}{d\Omega\,d(\nu/\nu_0)}
    \label{eq:wedge}
\end{equation}
with $\nu_0=1420\,$MHz and $n_{\rm pol} = 2$ the number of polarizations.  The effective area is $A_e = \pi D_e^2/4$, related to the physical area by an aperture efficiency, $\eta_a = 0.7$, such that $D_e^2 = \eta_a D^2$.  We take the field-of-view per pointing to be $\Omega_p(z) = (\lambda / D_e)^2$ -- see e.g.\ ref.~\cite{Chen19}.  In a spatially flat model
\begin{equation}
  \frac{d^2V}{d\Omega\,d(\nu/\nu_0)} = \chi^2\,\frac{d\chi}{dz}\,\nu_0\,\frac{dz}{d\nu}
  = \chi^2\,\frac{c\,(1+z)^2}{H(z)} \quad .
\end{equation}
The system temperature is the sum of amplifier noise, sky and ground temperatures, $T_{\rm sys} = T_{\rm ampl} + T_{\rm sky} + T_{\rm ground}$, with \cite{CVDE-21cm}
\begin{equation}
    T_{\rm ampl} = 50\eta_c^{-1} \,{\rm K}, \;
    T_{\rm sky} = 2.7\,{\rm K} + 25\,{\rm K}  \left(\frac{\nu_{\rm obs}}{400\,{\rm MHz}} \right)^{-2.75}, \;
    T_{\rm ground} = \frac{1-\eta_c}{\eta_c}300\,{\rm K}   .
\end{equation}
and we assumed a 10\% loss of power in the amplifier $1-\eta_c=0.1$.
Shot noise from the finite number of halos and galaxies hosting HI  \cite{Castorina17,Padmanabhan15,Padmanabhan17},
\begin{equation}
\label{eq:PSN}
  P_{\rm sn}  = \frac{\int_0^\infty n(M_h;z) M_{\HI}^2(M_h;z)  \,\mathrm{d}M_h}{[\int_0^\infty n(M_h;z) M_{\HI}(M_h;z)  \,\mathrm{d}M_h]^2} \; ,
\end{equation}
with $n(M_h;z)$ the halo mass function and $M_{\HI}(M_h;z)$ the HI mass as a function of halo mass and redshift, will also introduce additional power into these visibility data. The shot noise level depends on the poorly known small scale physics of neutral hydrogen on halo and galactic scales, which should then be marginalized over \cite{Padmanabhan2018}. In this work we marginalize over the shot noise amplitude assuming a fiducial value computed using Eq.~\ref{eq:PSN} and the HI model in ref.~\cite{Castorina17}. This is what is done in a standard analysis of galaxy surveys and is closer to what a data analysis of 21-cm data might look like.

The full observed 21-cm signal is given by the sum of the cosmological signal, proportional to the HI power spectrum, and the noise terms,
\begin{equation}
    P_{21}(k,\mu) = \bar{T}_b^2[P_{\HI}(k,\mu) + P_{\rm sn}] + P_{\rm th}.
\end{equation}
The above expression shows that in the linear regime the brightness temperature $\bar{T}_b(z)$ is completely degenerate with the amplitude of cosmological fluctuations $\sigma_8(z)$.

The range of scales which can be measured in a 21-cm inteferometer depends upon the baseline configuration (i.e.~the size and spacing of the receivers that will be correlated) and assumptions about how well foregrounds can be subtracted \cite{Furlanetto06,Shaw14,Pober15,Seo16,Cohn16}.
The precise range of scales affected by foregrounds is currently a source of debate.  We include these complexities by restricting the range of the $k_\perp - k_\parallel$ plane we include in our forecast.  There are two regions of this plane we could lose to foreground removal.  The first is low $k_\parallel$ modes, i.e.~modes close to transverse to the line-of-sight, which are swamped by the (large amplitude but spectrally smooth) foregrounds. For an optimistic scenario, we follow \cite{Shaw14,Shaw15} and assume only modes with $k_\parallel < 0.01 \, h \text{Mpc}^{-1}$ are unusable. Our pessimistic scenario raises this to $0.1 \, h \text{Mpc}^{-1}$, as suggested in Ref.~\cite{Pober15}.  In addition to low $k_\parallel$, imperfect calibration of the instrument leads to leakage of foreground information into higher $k_\parallel$ modes.  This is usually phrased in terms of a foreground ``wedge'' which renders modes with low $k_\parallel/k_\perp$ unusable \cite{Datta10,Morales12,Parsons12,Shaw14,Shaw15,Liu14,Pober15,Seo16,Cohn16,Morales19}.
Our optimistic choice will be the `primary beam wedge' defined with $\theta_{FOV}\approx 1.22\lambda/2D$, where the factor of two gives an approximate conversion between the first null of the Airy disk and its FWHM.  We shall contrast this ``optimistic'' assumption with the ``pessimistic'' case $\theta_{FOV}\approx 3\times 1.22\lambda/2D$.  The wedge could potentially be even larger, possibly up the ``horizon wedge'' (see Refs.\ above).  We do not consider this possibility, as it renders the 21-cm experiment ineffective as a large-scale structure probe.

In this work we consider two instruments, one under construction, HIRAX, and the Stage {\sc ii} 21-cm experiment proposed in \cite{CVDE-21cm}. We expect results for CHIME to be qualitatively similar to the ones for HIRAX.

HIRAX \cite{HIRAX} is a $32\times32$ array of six meters, $D=6\,$m, fully illuminated dishes under construction in South Africa, observing at frequency corresponding to $0.8<z<2.5$.
The Stage {\sc ii} 21-cm experiment suggested by Ref.~\cite{CVDE-21cm} is close to be a scaled version of HIRAX, consisting of a compact, square array of $256\times 256$ six meter dishes, observing half the sky in the redshift range $2<z<6$.
We shall take these two instruments as indicative of current and possible future surveys.
The details of the noise model we will assume for both instruments can be found in Ref.~\cite{Chen19}.

\begin{figure}
    \centering
     \resizebox{.485\columnwidth}{!}{\includegraphics{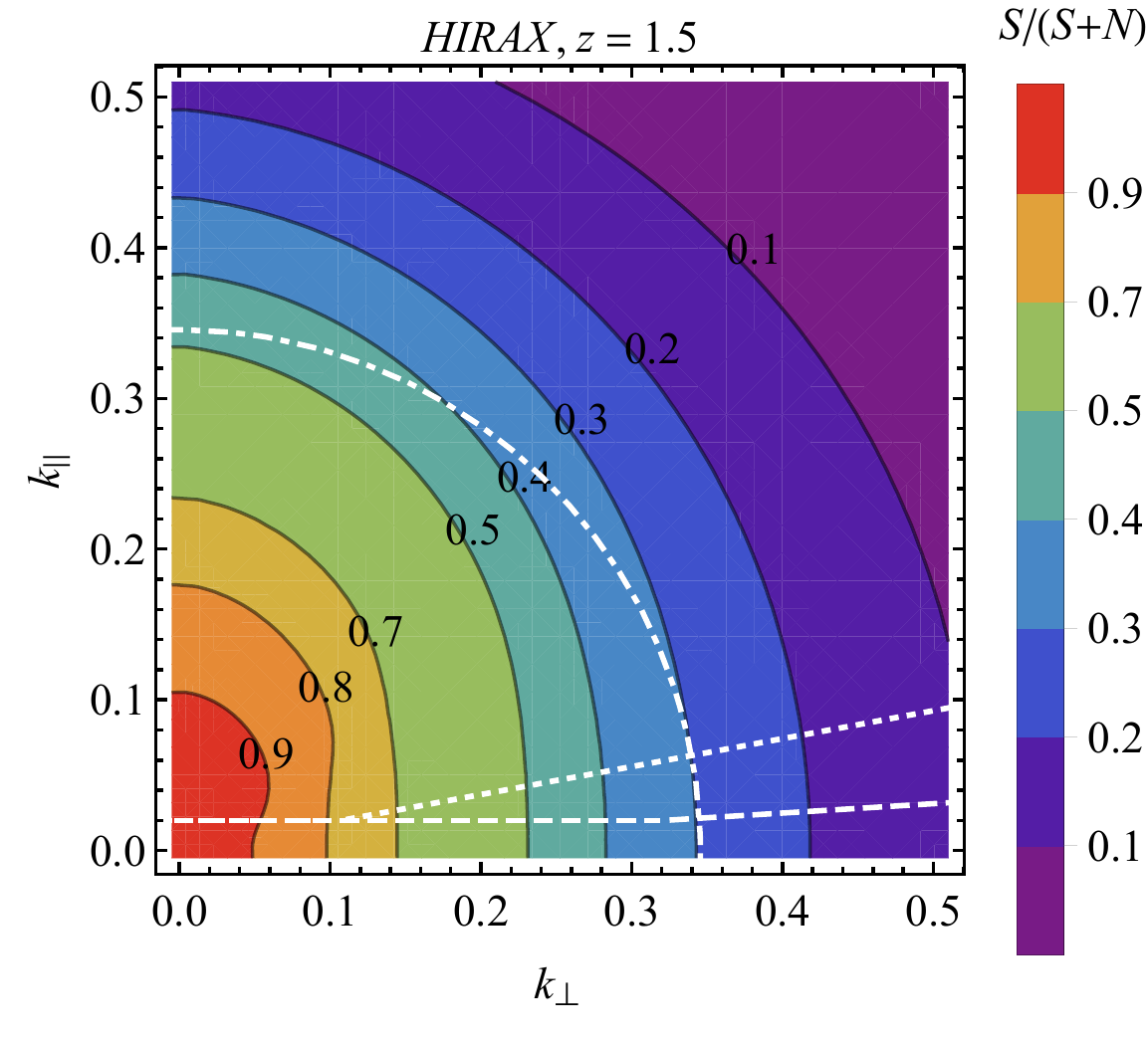}}
     \resizebox{.485\columnwidth}{!}{\includegraphics{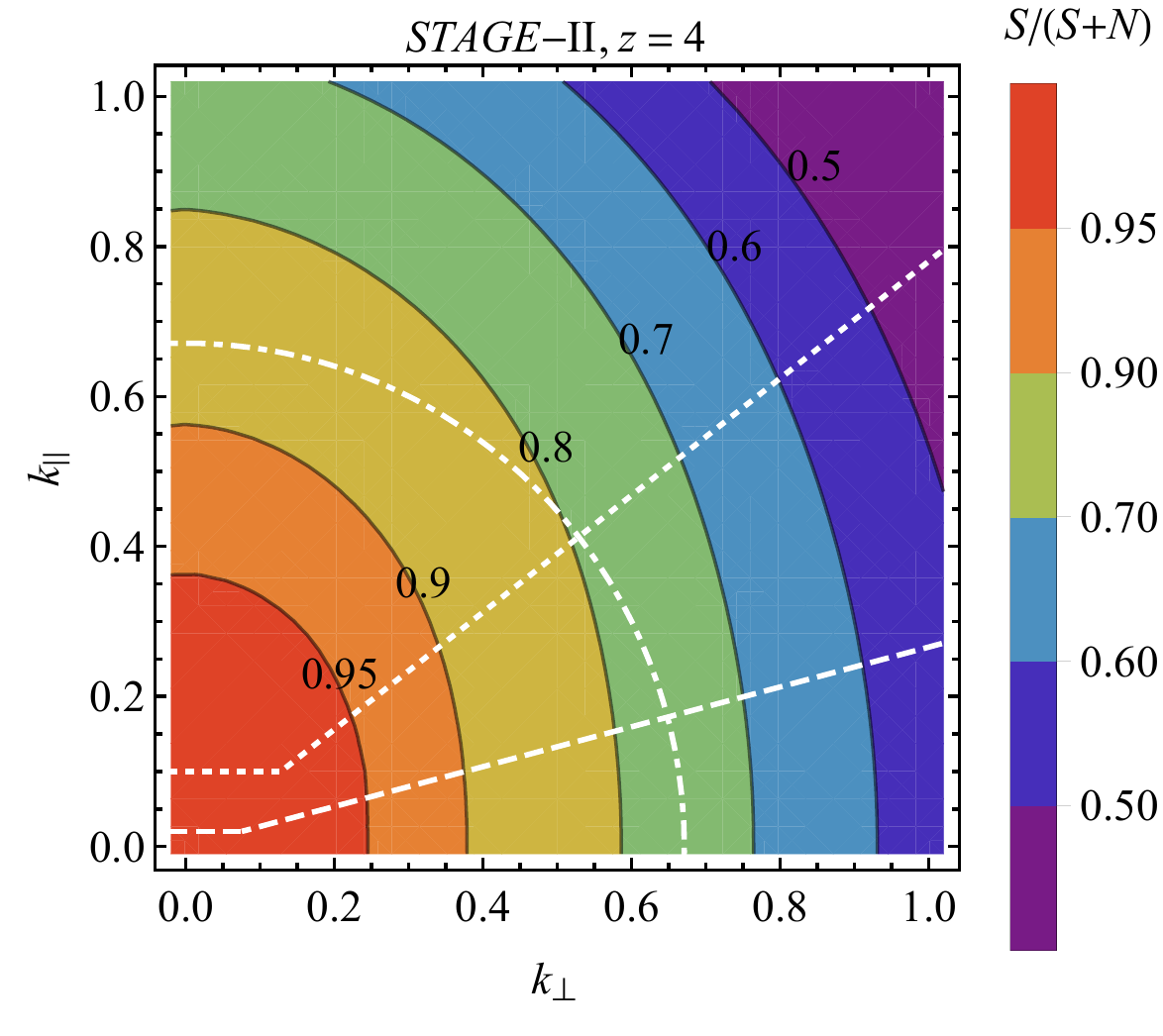}}
    \caption{The fraction of the total power in HI, $P_{\rm HI}/(P_{\rm HI}+P_{\rm sn}+P_{\rm th})$, at $z=1.5$ (left) and $z=4$ (right) for two proposed 21-cm instruments: HIRAX (left; \cite{HIRAX}) and Stage {\sc ii} (right; \cite{CVDE-21cm}).  The color scale shows the signal fraction while the white lines show the range of scales lost to foregrounds in the optimistic (dashed) and pessimistic (dotted) scenarios.  Only modes above and to the left of those lines are recoverable.  The white dot-dashed line shows $k_{\rm nl}$, which we take to be the upper limit to the scales that can be reliably modeled with perturbation theory.}
    \label{fig:wedge}
\end{figure}

To summarize, \fig{fig:wedge} shows the cosmological signal in units of the total power, as a function of $k_\parallel$ and $k_\perp$, for our benchmark 21-cm instruments: HIRAX (at $z=1.5$) in the left panel, and Stage-II (at $z=4$) in the right panel. 
The dashed and dotted lines represent our two choice for foreground removal\footnote{For HIRAX, a cut at $k_{\parallel}^{\text min} = 0.1\,\kMpc$ dominates over the wedge and it is therefore not shown in the plot.}.
The dot-dashed lines corresponds to the non-linear scale, $k_{\text{nl}}$, \ie the smallest scales at which perturbative calculations of the sort discussed in this paper can be trusted\cite{Carlson13},
\begin{equation}
   k_{\text{nl}}^{-2}\equiv  \Sigma^2 = \frac{1}{6\pi^2} \int P(k) \mathrm{d}k
   \quad .
    \label{eq:sigma}
\end{equation}

For HIRAX, almost all the modes we could in principle exploit using perturbation theory have signal-to-noise per mode larger than one, \ie $S/(S+N)>0.5$, which makes it an ideal survey for cosmological analysis. We see that the foregrounds have little impact, and most of the information remains available (as discussed above this only applies if the wedge does not extend all the way to the horizon). 
The uncertainty in a measurement of the anisotropic power spectrum is larger along the $k_\perp$ direction, as expected since the thermal noise only affects perpendicular modes. In traditional galaxy surveys the signal in high $k_{\parallel}$ modes is usually suppressed by random motions of the galaxies within their host halo, a non-perturbative effect known as Fingers-of-God (FOG). This is less true for the low mass halos where neutral hydrogen resides at high redshift \cite{VN18}, making 21-cm surveys also appealing from this perspective.
We explicitly checked that removing FoG doesn't change any of our conclusions\ as they only become important at $k_{||}>1\, \kMpc$ \cite{VN18,HVModi}.

The Stage-II case at $z=4$ is shown in the right. At high redshift the non linear scale has moved to larger wavenumbers compared to HIRAX, but still all the modes with $k<k_{\text{nl}}$ can be measured with very high signal to noise, with less than 20\% loss of information compared to a noiseless case. This plot suggests that even modes with $k\simeq1 \,\kMpc$ can be measured quite accurately, such that further gains can be obtained if numerical methods can be developed to supersede analytical models with parameterized bias.
The figure also informs us that, within the perturbative regime, a smaller array could also suffice to measure the growth of structure using RSD. This possibility is discussed in Appendix \ref{sec:miniS2}. 
At higher redshift the wedge has a larger impact and effectively cuts out half of the plane, but still lots of modes remain available for cosmological analyses.

\section{Forecasting methodology}
\label{sec:fisher}

We investigate the constraining power of 21-cm surveys through a Fisher matrix formalism \citep{Tegmark,White09}. We work at the level of the fields and assume the covariance matrix is diagonal in wavenumber.  We can then construct a Fisher matrix for the parameters $\theta_i = \{ \bar{T}, \sigma_8, b_1, b_2,P_{\rm sn}\}$, given by \cite{Tegmark97}
\begin{equation}
    F_{ij} = V_{\rm obs} \int \frac{d^3k}{(2\pi)^3}\ \frac{1}{2} \text{Tr}
    \left[ C^{-1} \frac{\partial C}{\partial \theta_i}
           C^{-1} \frac{\partial C}{\partial \theta_j} \right]
           = \frac{V_{\rm obs}}{2}\int_0^{k_{\rm max}}\frac{k^2\,dk}{2\pi^2}
           \int_0^1 d\mu\ \frac{\partial_i P\ \partial_j P}{P_{21}^2(k,\mu)}\,,
\end{equation}
where $i$, $j$, run over the parameters and $V_{\rm obs}$ is comoving volume of the interferometric survey.  We take $P_{HI}=\sum_{\ell=0}^4 P_{HI,\ell}(k)\mathcal{L}_\ell(\mu)$, with $\mathcal{L}_\ell$ the Legendre polynomial of order $\ell$.  For most of the parameters the derivatives of $P_{HI,\ell}$ in Eq.~(\ref{eqn:Pk_ZEFT}) can be performed analytically, and we do the $\sigma_8$ derivative using centered finite difference.

\begin{figure}
    \centering
    \resizebox{\columnwidth}{!}{\includegraphics{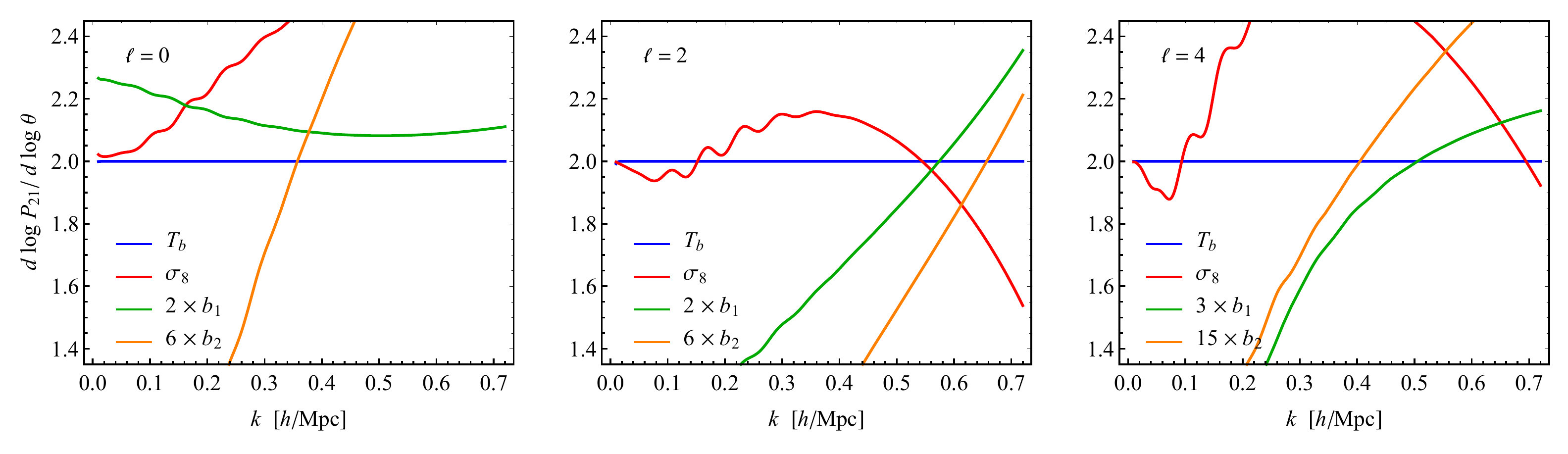}}
    \caption{The parameter derivatives of the monopole ($\ell=0$; left), quadrupole ($\ell=2$; middle) and hexadecapole ($\ell=4$; right) moments of the redshift-space power spectrum at $z=4$ for our fiducial model.  In all cases $d\ln P_\ell/d\ln\bar{T}$ is simply 2, and $d\ln P_\ell/d\ln(\sigma_8)$ is close to 2 at low $k$.  The scale-dependent bias term, $b_2$, is sub-dominant on the scales plotted (we have multiplied the plotted lines by large factors for plotting purposes).}
    \label{fig:derivs}
\end{figure}

\fig{fig:derivs} shows the (log) derivatives with respect to $\bar{T}$, $\sigma_8$ and the two bias parameters for the multipoles at $z=4$.  The log-derivative with respect to $\bar{T}$ is scale-independent and equal to 2.  At low $k$ the $\sigma_8$ derivative is also close to 2, and shows only a mild scale dependence until $k\simeq 0.1\,h\,{\rm Mpc}^{-1}$.  This is a manifestation of the degeneracy\footnote{The degeneracy is also sometimes written as being between $\bar{T}$ and $f\sigma_8$, but we assume $f$ is known.} between $\bar{T}$ and $\sigma_8$ in linear theory.  The small oscillations correspond to the baryon acoustic oscillations, which is damped by different amounts in models with different normalization.
At very low $k$ the $b_1$ derivative is also nearly independent of scale for the monopole, however it exhibits strong scale-dependence for $\ell=4$ and non-trivial dependence even for $\ell=2$.  In linear theory and with constant large-scale bias ($b=1+b_1$; \cite{Mat08}) we have $P_{HI}(k,\mu)=[(1+b_1)+f\mu^2]^2P_L(k)$ so $d\ln P_4/d\ln b_1=0$ and $d\ln P_2/d\ln b_1=(1+b_1)^{-1}$ while $d\ln P_0/d\ln b_1\approx 1$ for our fiducial parameters.  For $\bar{T}$ and $\sigma_8$ the (log) derivatives are both 2 and all of these (log) derivatives are $k$-independent.  Note, however, that in the Zeldovich approximation there is more structure to the derivatives due to the improved modeling of non-linear effects and the non-multiplicative effects of biasing and this allows us to break some degeneracies. Similar results hold at other redshifts.

The smaller scales included in our analysis are dictated by the non-linear scale, and we shall choose values between $k_{\rm max}/ k_{\rm nl}=0.5$ (very conservative) and $k_{\rm max}/ k_{\rm nl}=1$ (aggressive), to understand the sensitivity of our predictions to this quantity.

As described above the particularities of the 21-cm signal impose additional integration limits: foreground removal sets a lower bound on $k_\parallel$ while the wedge provides a lower bound on $\mu$ for usable $k$-modes. In what follows we will investigate the ``optimistic'' and ``pessimistic'' scenarios described above (and shown in Fig.~\ref{fig:wedge}).

\section{Results}
\label{sec:results}
\begin{figure}
    \centering
    \includegraphics[width=0.485\textwidth]{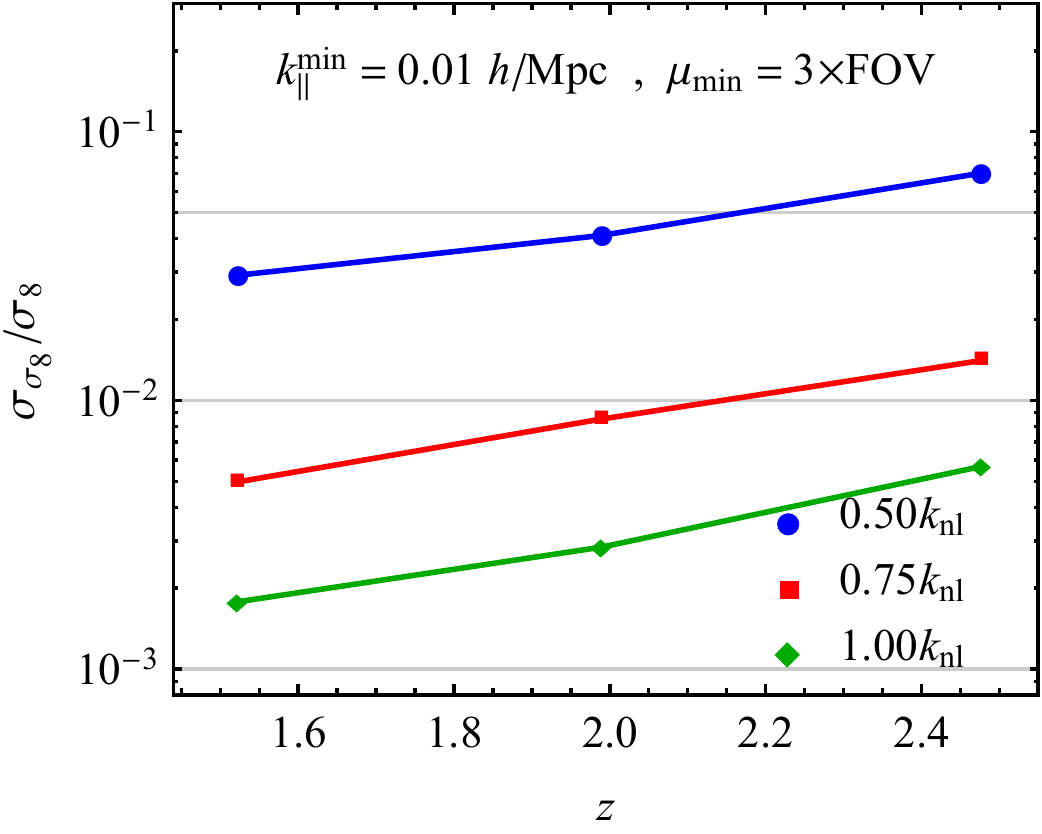}
    \includegraphics[width=0.485\textwidth]{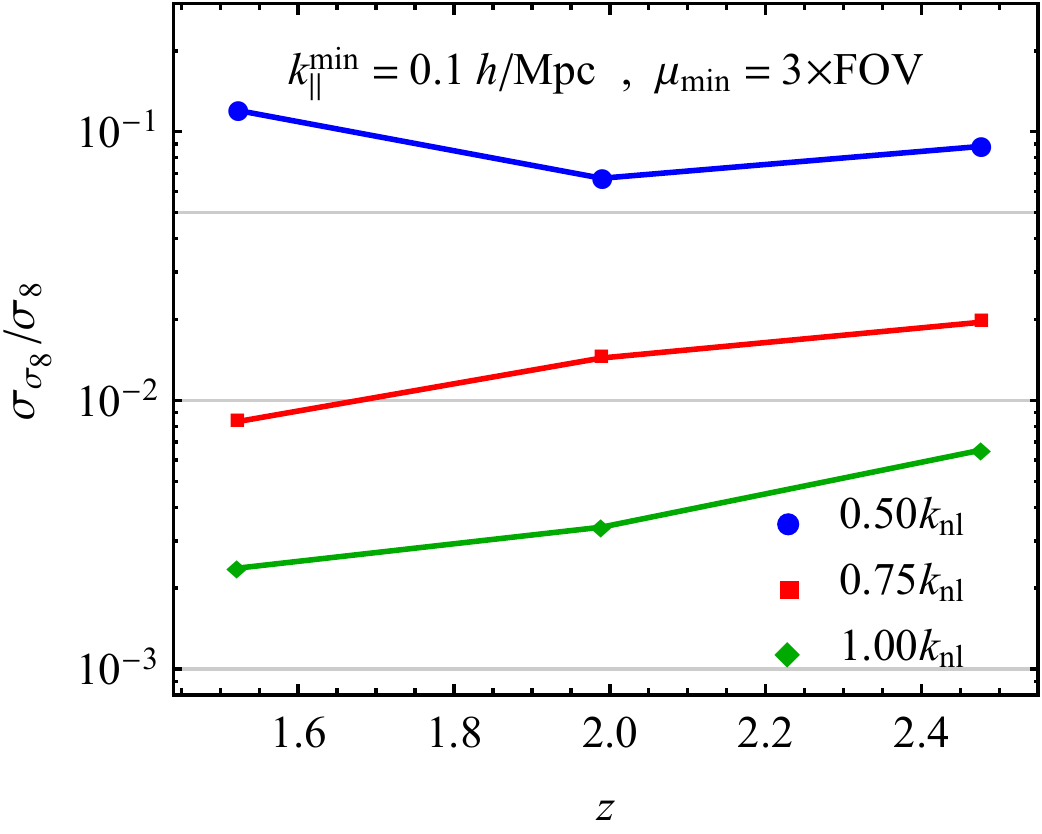}
    \caption{Forecast fractional error on the power spectrum normalization, $\sigma_8$, from HIRAX.  The left panel shows an optimistic scenario with good foreground cleaning while the right panel shows a pessimistic scenario.  Blue lines show constraints from $k_{\rm max}=0.50\,k_{\rm nl}$ while red lines show $k_{\rm max}=0.75\,k_{\rm nl}$ and green those for $k_{\rm max}=1.00\,k_{\rm nl}$.  These span the range of likely possibilities.  The non-linear information and the scale-dependent bias have dramatically broken the $\bar{T}_b-b-\sigma_8$ degeneracy which limits growth-of-structure measurements.}
    \label{fig:HIRAX}
\end{figure}

\fig{fig:HIRAX} shows the forecasted constraint on $\sigma_8(z)$ from HIRAX in the redshift range $1.5<z<2.5$. We marginalize over $b_1\,,b_2$ and $T_b$ in each redshift bin of width $\Delta z=0.5$ and assume as sky fraction $f_{\rm sky}=0.5$. 
The left panel considers the case of foreground removal down to $k_{\parallel}^{\rm min} = 0.01 \,\kMpc$, and wedge extending to 3 times the size of the primary beam. Different choices of $k_{\rm max}$ are shown with different colors. For the very conservative value of $k_{\rm max}$ (the blue line in \fig{fig:HIRAX}) we obtain a few percent constraint on $\sigma_8(z)$ at $z=1.5$ and a 6.5\% constraint at $z=2.5$. This indicates that even with a short lever arm in $k$, due to small value of $k_{\rm max}$, we are still able to break the degeneracy between cosmology and $\bar{T_b}$ (and bias) present in linear theory.
Pushing to $k_{\rm max}=0.75 k_{\rm nl}$, which we can probably reach with current perturbative methods \cite{Carlson13,Vlah16,Foreman16,Modi17}, the combination of non-linear structure formation and more complex bias has broken the degeneracy enough that the errors are $\simeq1\%$ or lower. The case of $k_{\rm max}=k_{\rm nl}$ assumes, optimistically, that the modeling could be extended to the non-linear scale without the need to introduce further bias or EFT parameters.
This level of constraining power is beyond the reach of any other planned LSS survey, and are made possible by the high signal-to-noise per mode of HIRAX in the mildly non-linear regime of structure formation.
The right panel shows the impact of removing more modes due to the presence of foregrounds, $k_{\parallel}^{\rm min} = 0.1\,\kMpc$. The conservative cut is the most affected since we have discarded a lot of the available modes, \eg at $z=1.5$ the non linear scale is $k_{\rm nl} = 0.34\,\kMpc$. The other two values of $k_{\rm max}$ instead still deliver percent accurate measurements of the amplitude of cosmological fluctuations. 
In all cases it was crucial to include the $\mu$-dependence of the signal to break the degeneracies. This can be clearly seen in \fig{fig:derivs}: the different multipoles have different response to bias parameters, $\bar{T}_b$ and $\sigma_8$.
Had we artificially set the anisotropic dependence of the clustering to zero, and used only the monopole part of the power spectrum, we would have gotten much worse constraints, with $\sigma_{\sigma_8}>10\%$ independently of $k_{\rm max}$ and the foreground model.


\begin{figure}
    \centering
    \includegraphics[width=0.485\textwidth]{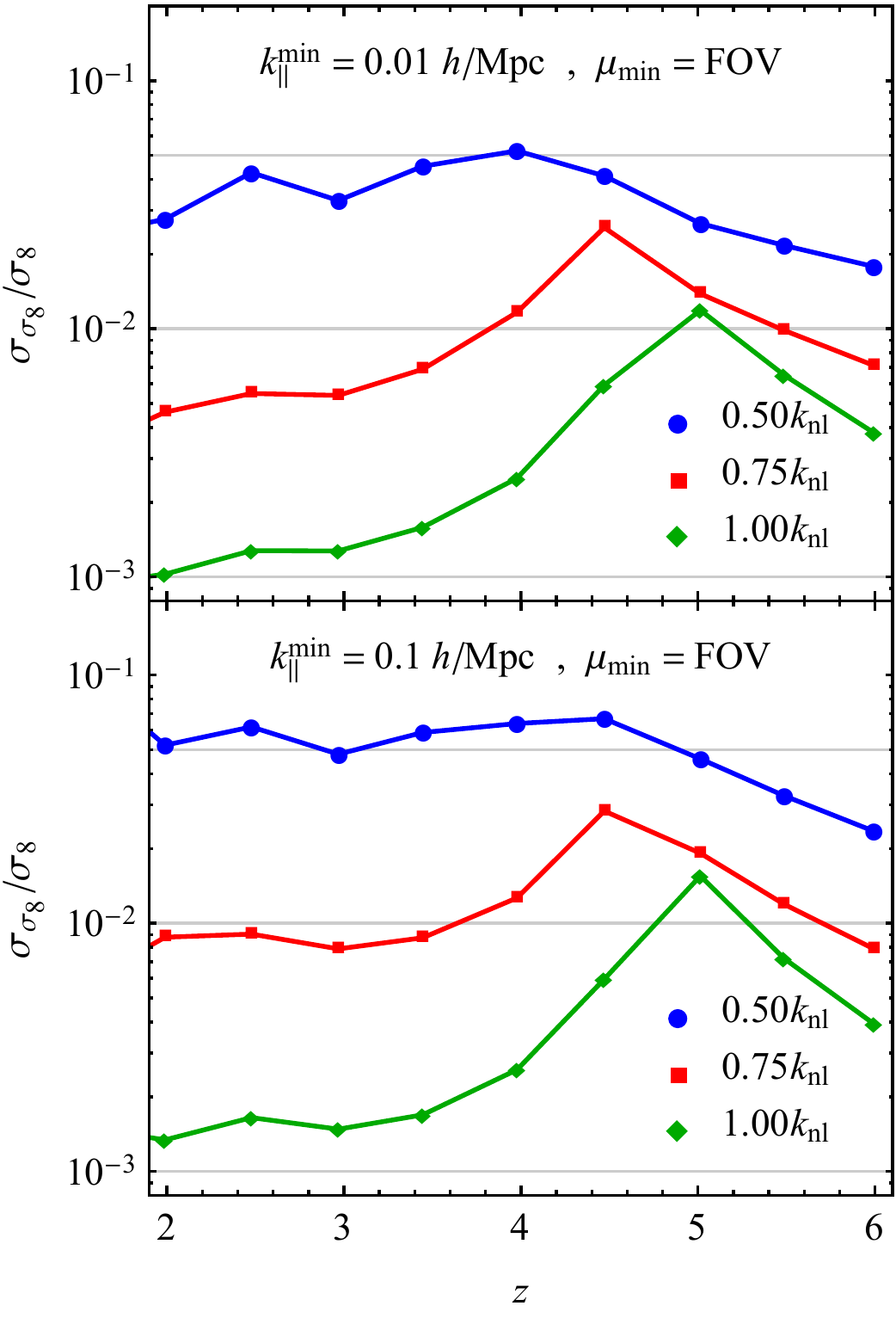}
    \hfill
    \includegraphics[width=0.485\textwidth]{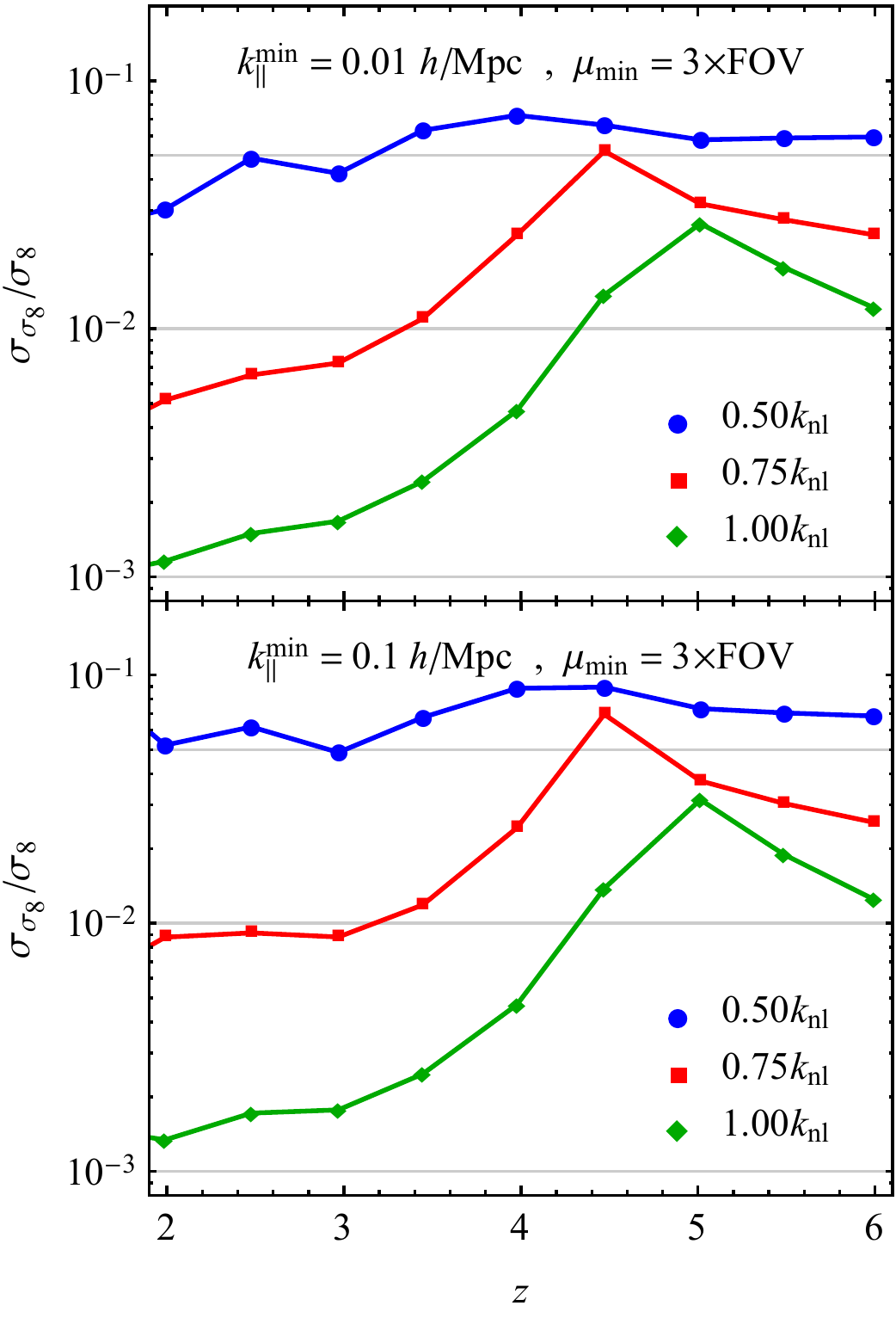}
    \caption{Forecast fractional error on the power spectrum normalization, $\sigma_8$, from a future 21-cm experiment.  The left panels shows an optimistic scenario with good foreground cleaning of the wedge, and two choices of $k_{||}^{\rm min}$, while the right panel shows a pessimistic scenario.  Blue lines show constraints from $k_{\rm max}=0.50\,k_{\rm nl}$ while red lines show $k_{\rm max}=0.75\,k_{\rm nl}$ and green those for $k_{\rm max}=1.00\,k_{\rm nl}$.  These span the range of likely possibilities.}
    \label{fig:S2}
\end{figure}

The constraining power of the Stage {\sc ii} instrument proposed in Ref.~\cite{CVDE-21cm}  is shown in \fig{fig:S2}. The left panels  presents two choices of $k_{\parallel}^{\rm min}$ for a foreground wedge that extends to the size of the primary beam, and the right panel the same two cuts in $k_{\parallel}$ but a three times larger wedge.
For reference the non-linear scales increases from $k_{\rm nl} = 0.4\,\kMpc$ at $z=2$ to $k_{\rm nl} = 0.9\,\kMpc$ at $z=6$.
As for the HIRAX case, the $k_{\rm max}=0.5\,k_{\rm nl}$ case is the most affected by the low $k_{\parallel}$ cut and the constraints on $\sigma_8(z)$ are approximately $5\%$. 
For higher values of $k_{\rm max}$ the constraining power of Stage {\sc ii} is impressive, and show the impact that continual improvements in modeling (with a combination of simulations and analytic approaches) could yield.  We regard the most likely situation as lying between the $k_{\rm max}=0.75\,k_{\rm nl}$ and $k_{\rm max}=0.50\,k_{\rm nl}$ cases.  While the data themselves may be fit to higher $k$, experience suggests that much of the information on those scales would serve to fix non-linear modeling and bias parameters (see e.g.~Ref.~\cite{Hand17} for an example at lower $z$).
In the range $2<z<5$, the dominant source of noise in the measurement is the shot-noise, which keeps increasing with redshift  and it is not compensated by a corresponding increment in the signal. This explains why the error bar on $\sigma_8(z)$ grows with redshift in this range.
At $z>5$ the thermal-noise becomes larger than the shot-noise, but the bias parameters have also increased substantially and allow us to further break the degeneracies  and reduce the errorbar on $\sigma_8(z)$. As emphasized in the previous section, the value of the HI bias at high redshift is extremely uncertain, therefore our results in this redshift range could quantitatively change depending on the the way HI traces the underlying matter distribution.

\begin{figure}
    \centering
    \resizebox{\textwidth}{!}{\includegraphics{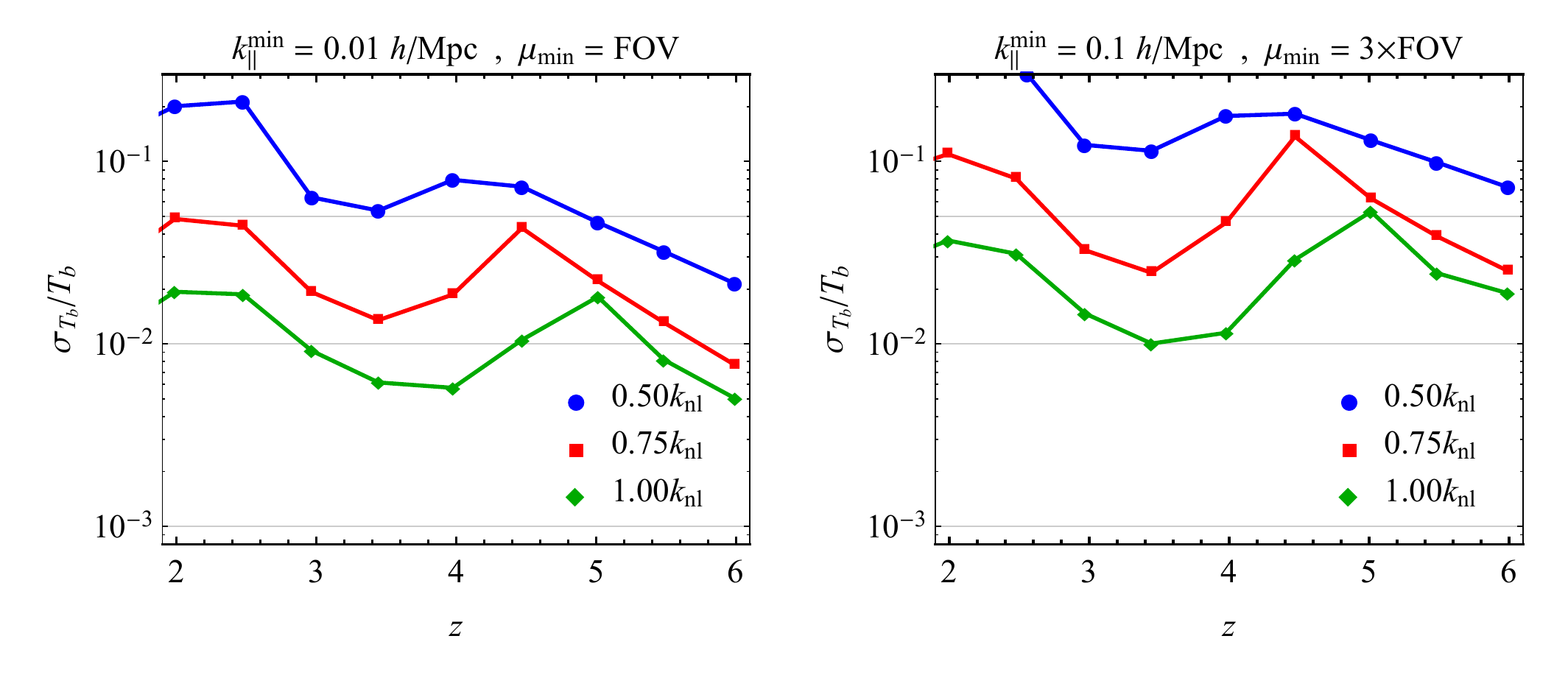}}
    \caption{Forecast fractional error on the mean brightness temperature, $\bar{T}_b$, from a future 21-cm experiment.  The left panel shows an optimistic scenario with good foreground cleaning while the right panel shows a pessimistic scenario. Color coding as in \fig{fig:S2}.}
    \label{fig:err_Tb}
\end{figure}

Breaking the $\bar{T}_b\mbox{-}\sigma_8\mbox{-}b_1$ not only allow us to measure the growth of structure over redshift, but also to constrain an important astrophysical parameter such as the cosmic abundance of neutral hydrogen, which is proportional to the mean brightness temperature.
\fig{fig:err_Tb} shows the fractional constraint on $\bar{T}_b(z)$ for the Stage {\sc ii} experiment. 
For the realistic choice of $k_{\rm max} =0.75 k_{\rm nl}$ we obtain few percent level constraints on the mean brightness of the 21 cm line in the redshift range $2<z<6$, which degrade by approximately a factor of 2 for the pessimistic foreground removal.
It would be hard to achieve similar percent level errorbars on the mean abundance of HI with measurements of the column density distribution function, as they are dominated by systematic uncertainties and the small number of high redshift quasars \cite{Crighton15}. We therefore conclude that an experiment like Stage {\sc ii} could also be a powerful astrophysical probe.

\section{Conclusions}
\label{sec:conclusions}

Intensity mapping surveys of HI using the 21 cm line at $z>1$ can, in principle, map vast expanses of space to constrain large-scale structure with unprecedented precision.
If foregrounds can be controlled and systematics well enough understood, such instruments would enable highly precise determinations of cosmological parameters.  Unfortunately the neutral hydrogen signal is modulated by the mean brightness temperature of the line, $\bar{T}_b$, which is only weakly constrained.  On the largest scales, this introduces a degeneracy between $\bar{T}_b$ and cosmological parameters such as the amplitude, $\sigma_8$, which cannot be broken without the use of external data.  However we have shown that mildly non-linear effects, in a regime which is still under theoretical control, effectively break this degeneracy.  At the high precision achievable on these scales with HIRAX or a Stage {\sc ii} 21-cm experiment these perturbative effects are sufficient to break the degeneracy without needing external data.

Our forecasts show that the breaking of the $\bar{T}_b-\sigma_8$ degeneracy by quasi-linear effects is sufficient to allow an both HIRAX and Stage {\sc ii} to make (sub-)percent level measurements of the growth of structure over a wide range of redshifts if the signal can be well modeled.  The precise constraint achieved depends upon the maximum $k$ to which the signal can be modeled, with $k_{\rm max}=k_{\rm nl}/2$ returning similar performance to cross-correlations \cite{Chen19} and $k_{\rm max}=k_{\rm nl}$ returning order of magnitude stronger constraints.  Being dominated by the higher $k$s our forecasts are relatively insensitive to assumptions about `the wedge' or low-$k_\parallel$ modes lost to foregrounds.  The constraints degrade to higher $z$, where the increase shot-noise and thermal noise weaken the power of a Stage {\sc ii} instrument.  It is also at these redshifts that the behavior of the 21-cm signal becomes increasingly important and unknown.
We have also shown the  Stage {\sc ii} could be quite useful in constraining, at the 5\% percent level or better, the mean HI abundance, a quantity of great astrophysical interest.

Our forecasts are done within the Zeldovich approximation, which allows for a treatment of complex bias and redshift space distortions relatively easily, while being more accurate than linear theory on quasi-linear scales and including more physics.  While not a full model of non-linear structure formation, we find it is quite accurate at high $z$ when compared to $N$-body simulations and captures much of the essential physics.  This makes it ideal for forecasting the reach of $21$-cm experiments where the significant unknowns about $21$-cm emission and the design of future instrumentation make the flexibility and speed of an analytic approach highly desirable.  We fully anticipate that fitting to any actual data would be done with a more complete model, or with an emulator based on numerical simulations. 
More generally, we would like to advocate the use of the Zeldovich approximation for forecasting in situations (like this one) where the matter field is quasi-linear on the scales of interest and the bias could potentially be complex.  The Zeldovich approximation is very stable and easy to compute and, while not a full non-linear model, it allows the incorporation of complex bias and matter-tracer decorrelation.  We believe it can be useful during exploratory work, to refine the region of interest before a more complex model (either from theory or numerical simulations) is used.

\section*{Acknowledgments}
EC and MW would like to thank the Cosmic Visions 21cm Collaboration and all its members for providing a great and stimulating scientific environment without which this work would have not been possible.
M.W.~is supported by the U.S.~Department of Energy and by NSF grant number 1713791.
This work made extensive use of the NASA Astrophysics Data System and of the {\tt astro-ph} preprint archive at {\tt arXiv.org}.

\vfill 

\appendix
\section{On the path to Stage {\sc ii}}
\label{sec:miniS2}
The  Stage {\sc ii} configuration proposed in Ref.~\cite{CVDE-21cm} consists of a $256\times256$ array of $6$ meters dishes. As we have seen in Section \ref{sec:signals} and \fig{fig:wedge}, this instrument has signal-to-noise much larger than one on any scale in the reach of perturbative methods.
This suggests that a smaller array could therefore deliver similar constraints, at a fraction of the cost. 

\begin{figure}
    \centering
    \includegraphics[width=0.49\textwidth]{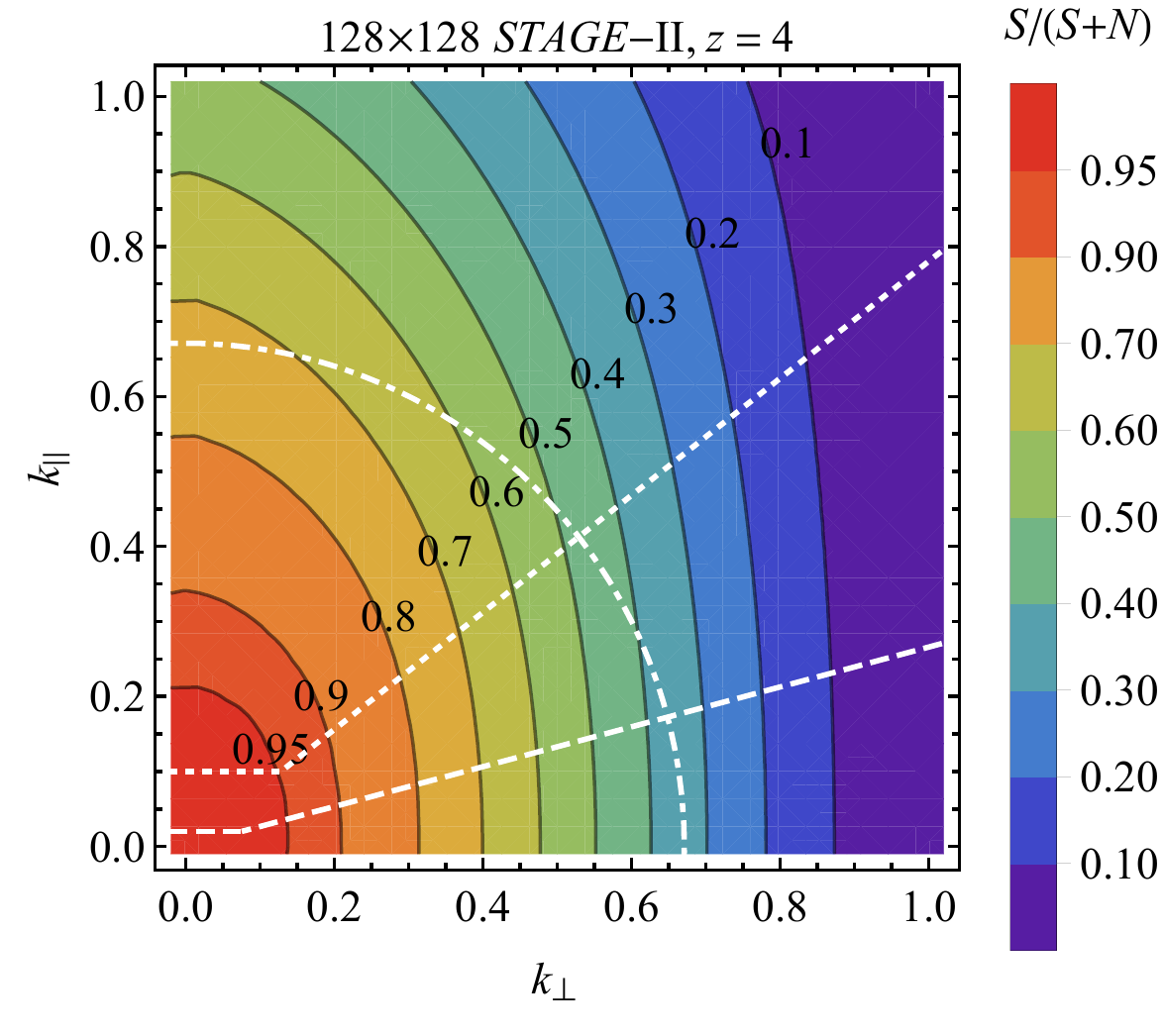}
    \includegraphics[width=0.48\textwidth]{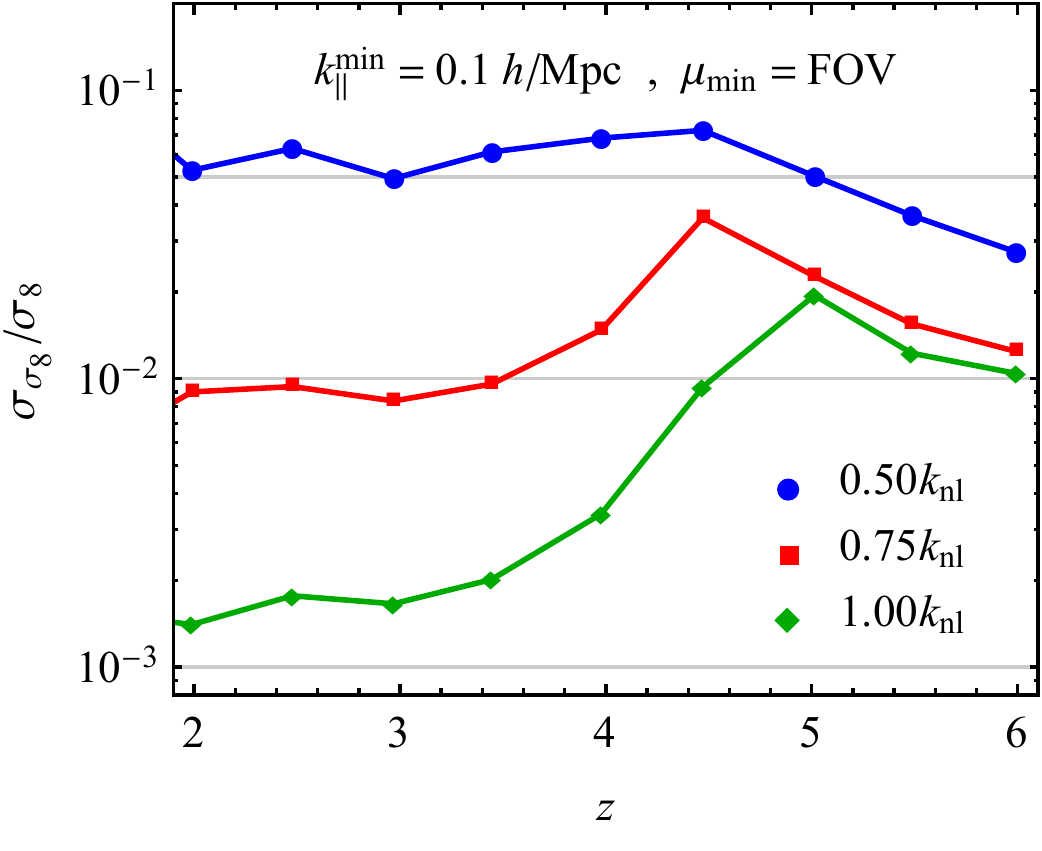}
    \caption{Left: The fraction of cosmological power in a measurement of the 21-cm power spectrum with a $128\times 128$ Stage {\sc ii} experiment at $z=4$. Right:  Fisher forecast of the amplitude of the power spectrum analogous to the bottom left panel of Fig.~\ref{fig:S2}.}
    \label{fig:miniS2}
\end{figure}

A proper optimization study of future 21-cm experiments is beyond the scope of this work, so in this section we limit ourselves to presenting the results of a Fisher analysis of a $128\times 128$ array of $6$ meter dishes, \ie 4 times smaller than the Stage {\sc ii}.
The left panel of \fig{fig:miniS2} shows the signal to noise in our rescaled version of Stage {\sc ii} at $z=4$, which should be compared with the full array shown in the right panel of \fig{fig:wedge}. Despite a reduced signal-to-noise compared to the larger experiment, a $128\times 128$ compact array is still able to measure most modes of interest for cosmology with high signal to noise.

The right panel in \fig{fig:miniS2} shows the corresponding forecast for $\sigma_8(z)$, assuming $k_{||}^{\rm min}=0.1\,\kMpc$. Compared to the full Stage {\sc ii}, the constraints at $z<4$ are almost indistinguishable, and only at the very end of the redshfit range the $128\times 128$ configuration performs substantially worse. The forecasted errorbars are still of the order of one percent, which indicates that at least for this science goal, a smaller array could be enough.

\bibliographystyle{JHEP}
\bibliography{main}

\end{document}